\documentclass[journal]{IEEEtran}
\ifCLASSINFOpdf
\else
\fi
\usepackage{bookmark}
\usepackage{graphicx}
\usepackage{url}
\usepackage{textcomp}
\usepackage[caption=false]{subfig}

\usepackage{csquotes}
\usepackage{booktabs} % For formal tables
\usepackage{framed}

\usepackage{tabularx}

\usepackage{url}
\usepackage{tabulary}
\usepackage{multirow}
\usepackage{enumitem}
\usepackage{hyperref}
\usepackage{hypcap}
\nocite{*}
\usepackage{lscape}
\usepackage[normalem]{ulem}
\useunder{\uline}{\ul}{}
\usepackage{longtable}
\usepackage{color, colortbl}
\usepackage{hyperref}

\definecolor{g}{gray}{0.925}
\usepackage{graphics}

\usepackage{booktabs}
\usepackage[flushleft]{threeparttable}
\usepackage{microtype}

\usepackage{placeins}

\usepackage{siunitx}  
\usepackage{textgreek}

\usepackage{flushend}

\usepackage{lscape}

\usepackage{rotating}

\begin{document}
\title{
The Double-Edged Sword of Diversity: \\How Diversity, Conflict, and Psychological Safety Impact  Software Teams
}

\author{Christiaan Verwijs and Daniel Russo
\thanks{C. Verwijs is with The Liberators, The Netherlands.}
\thanks{D. Russo is with the Department of Computer Science, Aalborg University, Denmark. Corresponding author. Email: daniel.russo@cs.aau.dk}
\thanks{Manuscript received Month 01, 2023; revised ....}}

% The paper headers
\markboth{IEEE Transactions on Software Engineering}%
{Verwijs \& Russo: The Double-Edged Sword of Diversity: \\How Diversity, Conflict, and Psychological Safety Impact Software Teams}

\maketitle

\begin{abstract}
Team diversity can be seen as a double-edged sword. It brings additional cognitive resources to teams at the risk of increased conflict. Few studies have investigated how different types of diversity impact software teams. This study views diversity through the lens of the \textit{categorization-elaboration model (CEM)}. We investigated how diversity in gender, age, role, and cultural background impacts team effectiveness and conflict, and how these associations are moderated by psychological safety. Our sample consisted of 1,118 participants from 161 teams and was analyzed with Covariance-Based Structural Equation Modeling (CB-SEM). We found a positive effect of age diversity on team effectiveness and gender diversity on relational conflict. Psychological safety contributed directly to effective teamwork and less conflict but did not moderate the diversity-effectiveness link. While our results are consistent with the CEM theory for age and gender diversity, other types of diversity did not yield similar results. We discuss several reasons for this, including curvilinear effects, moderators such as task interdependence, or the presence of a diversity mindset. 
With this paper, we argue that a dichotomous nature of diversity is oversimplified. Indeed, it is a complex relationship where context plays a pivotal role. A more nuanced understanding of diversity through the lens of theories, such as the CEM, may lead to more effective teamwork.
%Despite the complex relationship between diversity and effectiveness, much progress has been made through integrated models like the CEM. A deeper understanding of diversity through this lens will lead to more effective teamwork and equal opportunities in an increasingly diverse society.
\end{abstract}

\begin{IEEEkeywords}
software teams, agile, diversity, psychological safety, conflict
\end{IEEEkeywords}

\section{Introduction}
Teams are increasingly important to organizations. This is particularly relevant with the rise of Agile software methodologies, which 80\% of organizations now predominantly employ for their software teams~\cite{version1stateofagile}. Agile represents a collaborative, iteration-based, and human-oriented approach to product development~\cite{manifesto2001agile}. %It originated in response to the perceived shortfalls of plan-based approaches in the face of complex problems typical in product development~\cite{larman2004agile,highsmith2001agile}. Thus, ``at its core, agile project management is about managing the impact of complexity and uncertainty on a project''~\cite[p. 281]{dybaa2014agile}.%
Many scholars have attempted to identify the factors and characteristics that influence team effectiveness~\cite{hackman1976design,hackman1983normative}. One factor that has gained increased attention in recent decades is team diversity~\cite{patricio2022systematic}, also in software engineering specifically~\cite{silveira2019systematic}. \textit{Team diversity} is generally defined as heterogeneity in member attributes, such as age, gender, cultural background, tenure, role, or personality traits~\cite{williams1998demography}. 

While teams can be diverse on many attributes, most studies focus on demographic diversity (e.g., age, gender, cultural background) or informational diversity (e.g., professional role, education, experience).
Many researchers have theorized that diversity improves team performance~\cite{russo2020gender,conboy2009creativity,silveira2019systematic}. However, studies have provided mixed support. 
Investigations of how diversity impacts teams~\cite{tshetshema2020systematic,bowers2000member,patricio2022systematic,horwitz2007effects,conboy2009creativity,silveira2019systematic} generally show that the effects are not clear-cut, vary by type of diversity, and appear to be moderated by characteristics of the task, the team, and its environment. However, diversity may also negatively impact effectiveness through an increased conflict between members~\cite{patricio2022systematic,tshetshema2020systematic}. Several competing mechanisms and integrated models have been proposed to explain these conflicting results~\cite{horwitz2005compositional,van2004work}, which are discussed in Section~\ref{sec:related}.

Specifically for software engineering, Silveira \& Prikladnicki~\cite{silveira2019systematic} and Rodríguez-Pérez, Nadri \& Nagappan~\cite{rodriguez2021perceived} concluded from literature reviews that our understanding of diversity in such teams still needs to be improved. They found that most studies have only investigated gender diversity~\cite{silveira2019systematic} and argue that a broader exploration of how diversity impacts software engineering teams can be used to create better teams and better results. A conceptual way to look at this is through the lens of \textit{team effectiveness}. Hackman~\cite{hackman1976design} defined this as the degree to which the outcomes of a team satisfy the expectations of those they work for, as well as its members. Verwijs \& Russo~\cite{verwijs2023theory} recently operationalized team effectiveness for Agile software teams through stakeholder satisfaction and team morale. 

Studies have yet to explore how diversity affects software teams and their effectiveness. A more comprehensive examination is vital to understand how to design more effective teams and achieve better results. Henceforth, our research question (RQ) is:\\

RQ: \textit{How does diversity in software teams impact their effectiveness?}\\

To answer our research question, we performed a quantitative cross-sectional study with 1,118 team members representing 161 software teams. Covariance-Based Structural Equation Modeling (CB-SEM or SEM in short) was used to test how four types of diversity (gender, age, cultural background, and role) and one social moderator (psychological safety) interact to impact team effectiveness and conflict in teams. Only age diversity was positively associated with team effectiveness. Concerning relational conflict, only gender diversity showed a significant positive association. A replication package is also openly available on Zenodo to support secondary studies.

%Action - Structure
The rest of the paper is structured as follows. In Section~\ref{sec:related}, we review the related works of team diversity and how it impacts team outcomes. Subsequently, we clarify the research gap this study intends to address and develop relevant hypotheses in Section~\ref{sec:researchgap}. Section~\ref{sec:researchdesign} clarifies how we use quantitative methods and a survey study to test our hypotheses. The study results are reported in Section~\ref{sec:analysis}, followed by a comprehensive discussion of the results and their implications in Section~\ref{sec:discussion}. Finally, we conclude our paper outlining future research opportunities in Section~\ref{sec:conclusion}.

\section{Related work}
\label{sec:related}
Scholars from several disciplines have shown mixed results regarding how diversity impacts team effectiveness. We note that most of these studies have investigated \textit{team performance}. While team effectiveness is often used interchangeably with team performance in literature, including by Hackman~\cite{hackman1983normative}, they are not necessarily the same. Some definitions put more emphasis on speed and volume of output~\cite{lynn2000measuring}, whereas others emphasize  creativity~\cite{bodla2018diversity} or learning~\cite{kim2017effects}. However, most definitions share that teams are able to produce high-quality outcomes.

Tshetshema \& Chan conclude from a review of 35 studies that ``\textit{a negative relationship between [demographic diversity] and team performance is inferred as the most reported result.}''~\cite[p. 9]{tshetshema2020systematic}. However, they note that investigations of individual dimensions of diversity often show a positive effect on team performance, particularly gender and age. The complex relationship between diversity and performance is also recognized by Patrício \& Franco~\cite{patricio2022systematic}. They argue from a review of 80 studies that diversity has a dual impact on performance. One is positive through expanding perspectives, and the other is negative through increased conflict. Bowers, Pharmer \& Salas~\cite{bowers2000member} performed a meta-analysis of 13 empirical studies and found the effects of team diversity on team performance to be dependent on task complexity and difficulty instead. Their results suggest that teams that perform tasks of low complexity may benefit more from homogeneity, whereas teams that perform complex tasks benefit from higher diversity. Another meta-analysis of 30 empirical studies by Horwitz \& Horwitz~\cite{horwitz2007effects} found no significant effect of demographic diversity (age, gender, or cultural background) on team performance but did find a significant moderate effect of role diversity.

We now turn to investigations of individual dimensions of diversity in teams commonly studied by scholars.

\subsection{Diversity dimensions}
For \textbf{age diversity}, Tshetshema \& Chan~\cite{tshetshema2020systematic} found a positive effect on team performance in a review of empirical studies. However, a meta-analysis of 74 empirical studies by Schneid et al. ~\cite{schneid2016age} did not show a significant relationship, although modest differences occurred as a result of moderators like task complexity and team. Pesch, Bouncken \& Kraus~\cite{bouncken2015sme} attribute the positive effect of age diversity primarily to differences in tenure and work experience rather than age itself. They also note that this diversity is likely to increase tension and conflict in teams as members have to reconcile more diverse perspectives on completing tasks.

\textbf{Cultural diversity} is defined as heterogeneity in shared beliefs, norms and values~\cite{hui2017effects}. It is often operationalized through surface-level ethnic or national diversity~\cite{tshetshema2020systematic}. Tshetshema \& Chan~\cite{tshetshema2020systematic} inventoried studies that investigated the link between cultural diversity and team performance and inferred that a positive relationship is the most reported result. However, the relationship appears curvilinear: moderate cultural diversity is beneficial, but too little or too much adversely affects team performance~\cite{hui2017effects}.

Scholars define \textbf{gender diversity} as heterogeneity in the gender of team members. Most studies suggest a positive relationship with team performance~\cite{tshetshema2020systematic,russo2020gender}. Nevertheless, too much diversity may lead to increased conflict, particularly for complex tasks and high interdependence. Thus, Haas \& Hartmut~\cite{haas2010can} argue that gender diversity should be avoided in such environments.

\textbf{Role diversity} is another dimension of diversity in teams that is frequently studied. It represents the heterogeneity in the functional disciplines and roles members bring to a team~\cite{pelled1999exploring}. Agile software methodologies in particular emphasize the need for role diversity in teams in order to solve complex problems~\cite{larman2004agile,manifesto2001agile}. Empirical studies have shown mixed results, with some demonstrating positive effects and others negative~\cite{van2004work}. Pelled, Eisenhardt \& Xin~\cite{pelled1999exploring} found that role diversity increases conflict due to the integration of diverse perspectives, which positively influences task performance. Homberg \& Bui~\cite{homberg2013top} found no significant effect of role diversity on the performance of management teams in a meta-analysis of 53 empirical studies. Instead, they attribute the mixed findings to publication bias where those studies that don't report significant effects are published far less often than those that do. Horwitz \& Horwitz~\cite{horwitz2007effects} did find a modest effect on the quality of the work produced by teams in another meta-analysis, though not on performance.

The empirical link between diversity and team performance appears to be complicated. Several moderators have been found to strengthen the positive impact or dampen the negative impact, such as an inclusive team climate~\cite{bodla2018diversity}, task complexity and difficulty~\cite{bowers2000member,horwitz2007effects}, psychological safety~\cite{edmondson1999psychological}, management support~\cite{wickramasinghe2015diversity}, or time~\cite{steffens2012birds}.

\subsection{Diversity in software teams}
The importance of team diversity has also been recognized for software teams specifically~\cite{rodriguez2021perceived,larman2004agile}. The assumption is that diversity allows for a richer exploration of shared problems due to the availability of more perspectives~\cite{conboy2009creativity,silveira2019systematic}. This is particularly relevant to the complex problem-solving in software teams, which requires creativity and the application of diverse skill sets~\cite{highsmith2001agile}.
Several studies have investigated whether this assumption holds up in practice. Lee \& Xia~\cite{lee2010toward} used a mixed-methods approach to investigate how role diversity and team autonomy influence the ability of (Agile) software teams to deliver on budget, on time, and within scope. They found a significant positive effect of diversity in a survey study of 399 software projects and follow-up case studies. However, they found that role diversity improves the quality of solutions emerging from problem-solving in teams, but not speed. They also found evidence for the dual impact of diversity, where diversity also increases conflict.
Melo et. al.~\cite{melo2013interpretative} performed a multiple-case study of software teams in three large Brazilian software companies. Their results suggest that teams are more productive when there is diversity in the experience that members bring to the team.
Another study by Russo \& Stol~\cite{russo2020gender} surveyed 483 software engineers to investigate how personality and gender influence the productivity of software teams. Their results show that men and women typically bring different positive and negative traits to teams, and they argue that this explains some part of why mixed-gender teams perform better.
Rodríguez-Pérez, Nadri \& Nagappan~\cite{rodriguez2021perceived} conclude from a literature review that gender differences between developers contribute significantly to how they solve problems, debug issues, and work with others. The authors also note that gender diversity is most frequently studied, but much less is known about how other types of informational and demographic diversity affect software teams. A similar conclusion is reached by Silveira \& Prikladnicki~\cite{silveira2019systematic} in a review of the literature on diversity in (Agile) software teams. Thus, both groups of authors call for more research to guide decision-making on how to design better teams and generate better results.

\subsection{Theories and moderators of the diversity-effectiveness link}
Two mechanisms have been proposed by which diversity influences team effectiveness~\cite{horwitz2005compositional}. The \textit{similarity-attraction paradigm}~\cite{williams1998demography} derives from social psychology and social categorization to argue that similarity between members increases mutual attraction, integration, and communication, which in turn improves effectiveness. Diversity of members, on the other hand, results in more conflict and misunderstandings as people categorize themselves into different subgroups. Jehn~\cite{jehn1994enhancing,jehn1995multimethod} conceptualizes such conflicts as ``relationship conflict'' and distinguishes them from ``task conflict''. Where task conflict involves disagreement on how to proceed with the work at hand, relational conflict emerges as interpersonal friction from differences in values, political preferences, personal tastes, and interpersonal style. While relational conflict clearly negatively affects team effectiveness~\cite{jehn1995multimethod,simons2000task}, some level of task conflict is often thought to be useful as it encourages deeper information processing~\cite{carnevale1998social}. However, a meta-analysis by De Dreu \& Weingart~\cite{de2003task} does not support that distinction and suggests that even low-level task conflict is detrimental to team effectiveness in most cases.

Another mechanism that explains how diversity influences team effectiveness is \textit{cognitive resource diversity theory}. It derives from cognitive psychology. It treats teams as information processors where individuals process information and then elaborate and integrate it as a team~\cite{hinsz1997emerging}. In this conceptualization, diversity allows teams to bring varied cognitive resources to bear when information is processed individually and elaborated as a team, which allows a richer exploration of shared challenges. 

Thus, both mechanisms offer conflicting predictions about how diversity will impact team effectiveness. The former expects relational conflict to increase and effectiveness to decrease, whereas the latter expects effectiveness to increase. However, the evidence mentioned above does not consistently support one or the other. 
So the focus of academic inquiry has shifted toward identifying potential moderators that allow both mechanisms to be integrated~\cite{van2004work,van2007work,horwitz2007effects,horwitz2005compositional}.

One potential group of moderators concerns task characteristics, like complexity and interdependence~\cite{horwitz2005compositional}. In this view, homogeneity benefits low-complexity tasks with few inter-dependencies, whereas heterogeneity benefits more complex tasks with many inter-dependencies. This is primarily consistent with findings from meta-analyses of the diversity-effectiveness relationship~\cite{bowers2000member,schneid2016age}. However, other studies have found both positive and negative effects of task interdependence on the relationship between diversity and effectiveness~\cite{van2004work}.

Another potential moderator is psychological safety. Edmondson~\cite[p. 9]{edmondson1999psychological} defines it as ``a shared belief held by team members that the team is safe for interpersonal risk-taking''. Several studies have already shown that psychological safety contributes to more effective teamwork in software  teams~\cite{verwijs2023theory,moe2010teamwork,strode2022teamwork,hennel2021investigating}. However, psychological safety is also likely to moderate the relationship between diversity and team effectiveness. Diegmann \& Rosenkranz~\cite{diegmann2017team} theorize that psychological safety makes teams more resilient against the disruptive effect of high diversity, such as increased conflict, by providing a safe environment for members to elaborate task information. Similarly, Roberge \& Van Dick~\cite{roberge2010recognizing} expect that psychological safety also interacts with the salience of a collective identity. Diversity only contributes to higher team effectiveness when members feel safe and identify strongly with their team.

To date, few studies have empirically investigated the role of psychological safety as a moderator of the diversity-effectiveness association. Singh, Winkel \& Selvarajan~\cite{singh2013managing} found that employee performance was higher among members of diverse teams that also exhibited high psychological safety. However, this study was limited to one organization and only considered racial diversity. Furthermore, Kirkman et al. ~\cite{kirkman2013global} found that Communities of Practice (CoP) performed better when diversity was paired with high psychological safety. Virtual teams also experience fewer drawbacks from diversity when they can elaborate information in psychologically safe environments~\cite{gibson2006unpacking}. 

Van Knippenberg, De Dreu \& Homan~\cite{van2004work} have proposed the \textit{categorization-elaboration model (CEM)} to integrate the double-edged nature of diversity in teams and potential moderators. The CEM is the most comprehensive model of work group diversity and its moderators at the time of writing and has received broad empirical support~\cite{guillaume2017harnessing,kearney2009and,van2007work,van2005diversity}. It distinguishes between moderators related to the task, like difficulty, complexity, and efficacy,  and moderators related to the team and the social processes in it, like trust and commitment. Both groups of moderators influence the ability of teams to leverage the informational advantage offered through diversity, though in different ways. In the case of task moderators, complex and challenging tasks are more likely to elicit extensive information processing in members~\cite{horwitz2007effects,dahlin2005team}, which is consistent with \textit{cognitive resource diversity theory}. The motivation of teams with their task has also been shown to positively moderate the effects of diversity on information processing~\cite{scholten2007motivated}. Another potential task moderator is task interdependence, which is generally defined as the degree to which the completion of tasks requires collaboration by team members~\cite{stewart2000team}. Teams with low interdependence see less interaction and thus experience fewer opportunities to leverage the benefit of diverse cognitive resources. However, empirical studies have found positive and negative effects of task interdependence on the relationship between demographic and role diversity and team effectiveness~\cite{van2007work}. This suggests that the effect is either not linear or subject to other moderators. 

At the same time, the CEM also proposes a mechanism by which diversity can harm teamwork. As members grow less similar and bring different perspectives to teamwork, this diminishes effectiveness when the social context of a team encourages social categorization into subgroups and elicits negative inter-group biases and identity threat~\cite{van2007work,williams1998demography}. This loss of social integration creates more potential for relational conflict and negatively impacts the ability of teams to elaborate information effectively and reduces their effectiveness. However, social moderators like trust and psychological safety allow team members to integrate more effectively to bring diverse perspectives and information-processing together and elaborate on them, which is consistent with the \textit{similarity-attraction paradigm}. 

A strength of this integrated approach is that it may explain the conflicting results found in the literature. The different mechanisms behind both groups of moderators independently strengthen or diminish the ability of teams to leverage diversity and can work in concert or in opposition. Thus, the CEM broadens the discourse around team diversity from a one-dimensional approach where it is either a risk or an asset to one where it can be both simultaneously. Finally, the CEM has clear, practical implications for diversity management that aim to reduce in-group bias, strengthen social moderators, and match diversity with the nature of the task~\cite{van2004work}.

\subsection{Research Gap \& Hypotheses}
\label{sec:researchgap}
This study aims to address two related research gaps. The first is that we want to answer the call by Silveira \& Prikladnicki~\cite{silveira2019systematic} and Rodríguez-Pérez, Nadri \& Nagappan~\cite{rodriguez2021perceived} for more investigations into how diversity affects software teams, and not limited to only gender diversity. A more comprehensive examination is vital to understand how to design more effective teams and achieve better results. The second research gap is that we want to investigate diversity in software teams through the lens of the CEM theory and its opposing mechanisms. 

To answer our research question, we will now develop seven hypotheses we aim to test in this study. Our first hypothesis is that diversity contributes to the effectiveness of software teams. Because such teams collaborate on complex and interdependent tasks~\cite{highsmith2001agile,dybaa2014agile,larman2004agile}, they should benefit from the expanded cognitive resources allowed by heterogeneity in gender, age, cultural background, and role. This reflects one mechanism by which diversity influences team effectiveness and is in accordance with both \textit{cognitive resource diversity theory} and the CEM that integrates it.\vspace{0.5em}

Hypothesis 1 (H1). \textit{Software teams are more effective when they are more diverse in gender (H1a), age (H1b), cultural background (H1c), and role diversity (H1d)}.\vspace{0.5em}

Our second hypothesis concerns the second and opposing mechanism of diversity. That is, we expect that increased diversity also results in more relational conflict in teams. This hypothesis reflects a core consequence of the \textit{similarity-attraction paradigm} and the CEM that integrates it.\vspace{0.5em}

Hypothesis 2 (H2). \textit{Software teams experience more relational conflict when they are more diverse in gender (H2a), age (H2b), cultural background (H2c), and role diversity (H2d)}.\vspace{0.5em}

Furthermore, we hypothesize that the increased relational conflict, in turn, negatively impacts the effectiveness of teams. This is consistent with the outcome expected by the \textit{similarity-attraction paradigm} and the CEM that integrates it. \vspace{0.5em}

Hypothesis 3 (H3). \textit{Relational conflict reduces the effectiveness of software teams}.\vspace{0.5em}

Following existing literature~\cite{verwijs2023theory,moe2010teamwork,edmondson1999psychological,edmondson2014psychological}, we expect that psychological safety is a critical factor in enabling team effectiveness through four different processes. The first involves a direct effect where psychological safety makes teams more effective by creating more opportunities to openly elaborate information, reconcile conflicting viewpoints, and find creative solutions~\cite{verwijs2023theory,moe2010teamwork}.\vspace{0.5em}

Hypothesis 4 (H4). \textit{Psychological safety increases the effectiveness of software teams}.\vspace{0.5em}

In the second process, psychological safety decreases relational conflict in teams by providing more opportunities to air grievances and discuss the tension between members.\\

Hypothesis 5 (H5). \textit{Psychological safety reduces the amount of relational conflict in software teams}.\vspace{0.5em}

Concerning diversity, we expect that psychological safety is a social moderator of the association between diversity and team effectiveness. Consistent with the CEM and Diegmann \& Rosenkranz~\cite{diegmann2017team}, and as respectively our third and fourth processes, we anticipate as that psychological safety is a social moderator that creates an environment where diverse teams can more effectively elaborate task-related information \textit{and} and experience less relational conflict than less diverse teams.\vspace{0.5em}

Hypothesis 6 (H6). \textit{The relationship between diversity in gender (H6a), age (H6b), cultural background  (H6c), and role (H6d) on the one hand and team effectiveness on the other is moderated by psychological safety}. \vspace{0.5em}

Hypothesis 7 (H7). \textit{The relationship between diversity in gender (H7a), age (H7b), cultural background  (H7c) and role (H7d) on the one hand and relational conflict on the other is moderated by psychological safety}.\vspace{0.5em}

\begin{figure*}
\centering
\includegraphics[height=2.8in]{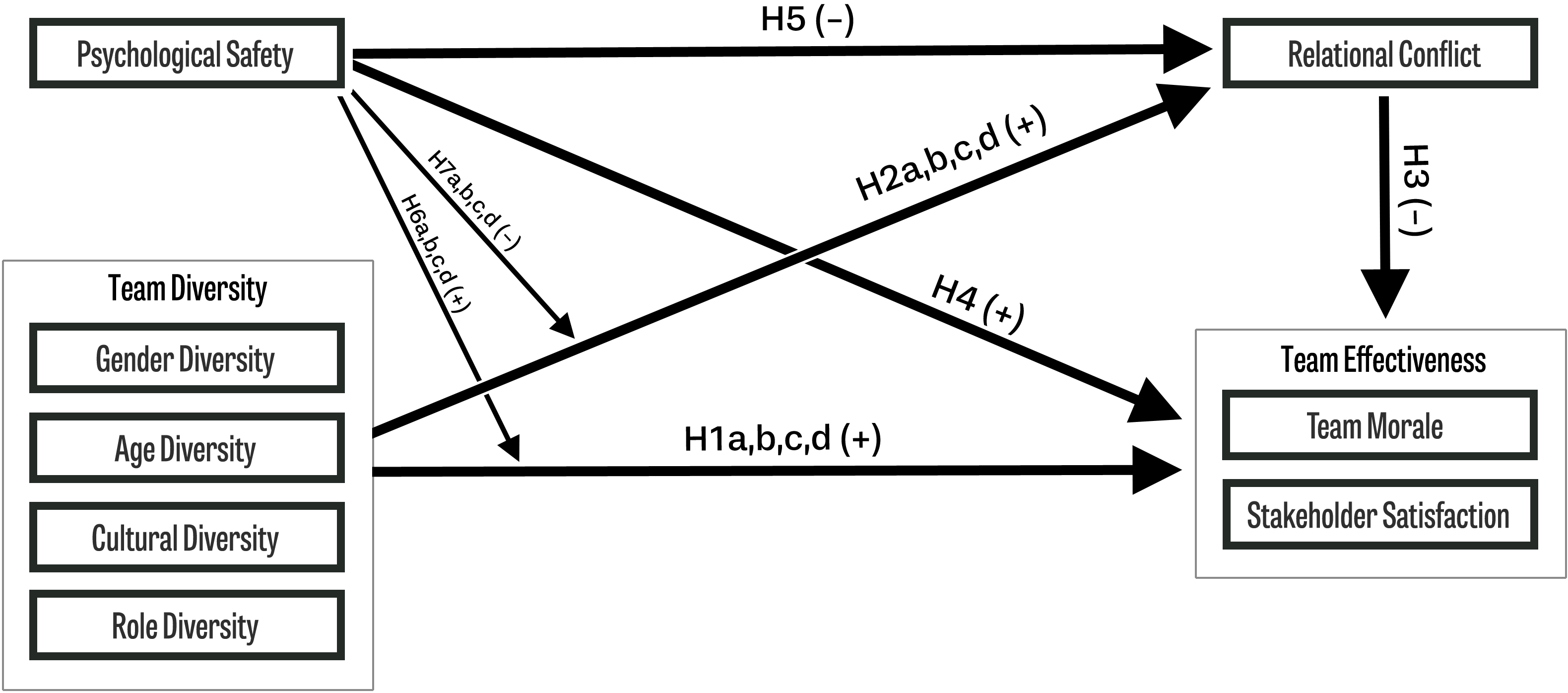}
\caption{Theoretical model and hypotheses. Sub-hypotheses are grouped, and control variables are omitted to retain visual clarity}
\label{fig:hypotheses}
\end{figure*}

\section{Research Design}
\label{sec:researchdesign}
We conducted a sample study with a sample of software teams to answer our research question. We used Covariance-Based Structural Equation Modeling (CB-SEM) to test our hypotheses (as visualized in Figure~\ref{fig:hypotheses}). This section discusses the sample (Sec.~\ref{sec:sample}), measurement instruments (Sec.~\ref{sec:measurements}), and method of analysis (Sec.~\ref{sec:analysis}).

\subsection{Participants}
\label{sec:sample}
We performed our data collection process through a customized online survey that was embedded in a larger survey that was part of an online tool for Agile teams~\footnote{The GDPR-compliant survey has been designed so that teams can self-assess their Agile development process. It is available at the following URL: www.scrumteamsurvey.org.}. A pilot study was first performed between July and September 2021 to identify improvements for the questionnaire. 256 teams participated. Two modifications were made. First, a scale for task-related conflict was removed because it was statistically indistinguishable from relational conflict in a Confirmatory Factor Analysis (CFA). Other studies have also reported this~\cite{de2003task}. Second, the worker councils of several participating organizations objected to an item that asked participants to identify their gender,  despite assurances that answers would remain anonymous. So the item was replaced with a team-level indication of gender diversity to overcome this objection (see Section ~\ref{sec:measurements} for more detail).

Data collection for the primary study was then performed between September 2021 and January 2022. A mix of purposive and respondent-driven non-probabilistic sampling strategies~\cite{baltes2022sampling} was used. While probabilistic strategies increase the likelihood of representative samples, they require knowledge of the true distribution of parameters in the population, which is typically not feasible for software engineering studies~\cite{russo2020gender}. Thus, we aimed to recruit experienced respondents by embedding the survey in a tool that is already used by many (Agile) software teams and promoted it across channels commonly visited by individuals interested in Agile and software development, including industry forums, blog articles, podcasts, and videos by influencers\footnote{Due to the opt-in nature of the survey, we were not able to compute a meaningful response rate. However, we were able to estimate that 65\% of the participants who started the survey completed it by sampling the number of uncompleted surveys from the total number of surveys before they were automatically removed after 14 days.}. To reduce sampling bias, we enabled respondents to also invite the members of their team to participate. This was encouraged by offering teams a report with anonymized team-level results for their team, along with helpful feedback. We were able to anonymously aggregate individual participants into teams as follows. Upon completion of a survey for a new team, each participant received a shareable link with a unique team key (a GUID) and instructions on how to invite the rest of their team. This participant then distributed that link through a channel of their choosing (e.g., email, Slack). We grouped all individuals by team key in the analyses.

In total, 1,827 members from 733 distinct software teams completed the survey in that period. Because the survey is public and accessible to anyone, we cannot properly calculate a response rate. Scholars have emphasized that public surveys are more susceptible to careless responses. So we applied several strategies outlined in literature~\cite{meade2012identifying} to reduce potential biases. First, we emphasized the anonymous nature of our data collection. Second, we encouraged honest answers by providing teams with a detailed team-level profile and relevant feedback for their team upon completion. Third, we removed 118 participants whose response patterns suggested careless responses. This consisted of participants who went through the survey too quickly to realistically read and answer the questions, thus preserving the robustness of the dataset. We followed a pragmatic cutoff described by Meade \& Craig~\cite{meade2012identifying} and removed all participants with a response time below the 5th percentile (6.87 minutes). We also included participants who answered fewer than half of the questions ($<20$), making them unsuitable for analyses. Finally, we retained only those teams (161) with at least 4 participating members to ensure a meaningful diversity measurement. The composition of our sample is shown in Table~\ref{tab:samplecomposition}.

Several variables in our model were measured at the individual level and aggregated to a higher (team) level in our analyses. Such aggregation is only reasonable when sufficient variance exists at the group level. The Intraclass Coefficients (ICC1 and ICC2) provide a measure for the proportion of group-level variance~\cite{hair2019multivariate}. We calculated this according to the procedure outlined by Van Mierlo et. al.~\cite{van2009composing} for all variables where individual-level measures were aggregated to the team level: psychological safety, relational conflict, team morale, and stakeholder satisfaction. Between 24\% to 32\% of the observed variation was attributable to the group level and indicated differences between teams rather than variation within teams (ICC1, $p < .001$). ICC1 values as low as 10\% can already indicate group-level relationships that do not emerge from individual-level analyses and warrant group-level analyses~\cite{thompson2012intraclass,bliese1998group,hair2019multivariate}. The group mean reliability (ICC2) indicates the agreement among members of each group and is typically interpreted as a reliability measure~\cite{nunnally1978overview,fleiss1986reliability}. Values closer to 1 indicate higher reliability, while those closer to 0 suggest lower reliability. For our study, it ranged between .68 and .77 and was deemed satisfactory. Given that values above .60 are generally considered satisfactory in many research contexts~\cite{nunnally1978overview,fleiss1986reliability}, our observed range underscores a significant level of agreement among team members on our various measures. This means that the ratings provided by individuals within the same team are relatively consistent, further justifying our aggregation approach. Taken together, the ICC1 and ICC2 both indicate robust group-level dynamics and warrant group-level analyses and aggregation.

We analyzed patterns in missing data at the individual level. Unless data is missing completely at random (MCAR), any patterns in missing data may bias the results of multivariate analysis~\cite{hair2019multivariate}. For this, we calculated Little's MCAR test. This is a Chi-Square test that compares the observed patterns of missing data with the patterns that would be expected from a process that results in random missing data~\cite{hair2019multivariate}. Our test showed that data wasn't completely missing at random  ($Chi^2=13,200.799$, $df=9,853$, $p<0.001$)~\cite{hair2019multivariate}. A closer inspection of patterns in the missing data revealed that the percentage of missing data was below 2\% for most items but slightly higher (up to 6.6\%) for four items that measured aspects relating to stakeholders. Overall, missing data remained below the recommended threshold of 10\%~\cite{hair2019multivariate}. No data was missing after we aggregated to the team level, so no imputation was performed.

We conducted an in-depth examination of missing data patterns at the individual level. It is well-established in the literature that when data is not missing entirely at random (MCAR), the presence of discernible missing data patterns can introduce potential bias into the results of multivariate analyses, as elucidated by Hair et al.\cite{hair2019multivariate}. We employed Little's MCAR test, a Chi-Square statistical procedure that juxtaposes the observed patterns of missing data against the patterns one would anticipate under the assumption of random data omissions\cite{hair2019multivariate}. Our examination of the data revealed a significant departure from the MCAR assumption, as evidenced by the test results ($Chi^2=1,013.462$, $df=778$, $p<0.001$)~\cite{hair2019multivariate}. Further analyses of the missing data patterns unveiled that the majority of items displayed missing data percentages lower than 3\%. However, we observed slightly higher rates for the four items related to stakeholder aspects of up to 6.6\%. Nevertheless, the percentage of missing data remained within the confines of the recommended threshold of 10\%~\cite{hair2019multivariate}. Because we aggregated data to the team level, at which no data was missing, we did not employ imputation procedures.

Finally, we performed a posthoc power analysis using G*Power~~\cite{faul2009statistical}, version 3.1.9. We determined that the sample size allows us to correctly capture medium effects ($f=.15$) with a statistical power of 96\% ($1-\beta= .96$). In other words, the probability of correctly rejecting the null hypothesis is 96\% given our sample. So we are confident that our sample is big enough to provide a reliable outcome.

\begin{table}[!t]
\centering
\caption{Composition of the sample}
\label{tab:samplecomposition}
\begin{tabular}{m{3cm}p{3cm}p{1.4cm}}
\toprule
\textbf{Variable} & \textbf{Category} & \textbf{N (\%)} \\ \midrule
Respondents &  & 1,118 \\
Teams &  & 161 \\
Respondents per team & 4-6 respondents & 82 (50.9\%) \\
 & 7-9 respondents & 64 (39.8\%) \\
 & 10+ respondents & 15 (9.3\%) \\
Product Type & Product for internal users & 89 (55.3\%) \\
 & Product for external users & 72 (44.7\%) \\
Scrum Team Size & 1-4 members & 1 (0.6\%) \\
 & 5-10 members & 129 (80.1\%) \\
 & 11-16 members & 26 (16.1\%) \\
 & \textgreater{}16 members & 5 (3.1\%) \\
Scrum Team Experience & Low & 5 (3.1\%) \\
 & Moderate & 75 (46.6\%) \\
 & High & 81 (50.3\%) \\
Organization Sector & Technology & 40 (24.8\%) \\
 & Financial & 29 (18\%) \\
 & Healthcare & 18 (11.2\%) \\
 & Other & 74 (46\%) \\
Organization Size & 1-50 employees & 10 (6.2\%) \\
 & 51-500 employees & 44 (27.3\%) \\
 & 501-5.000 employees & 50 (31.1\%) \\
 & \textgreater{}5.000 employees & 55 (34.2\%) \\
 & Not stated & 2 (1.2\%) \\
Role Diversity & Developer & 534 (47.8\%) \\
 & Scrum Master & 127 (11.4\%) \\
 & Product Owner & 104 (9.3\%) \\
 & Tester & 93 (8.3\%) \\
 & Analyst & 73 (6.5\%) \\
 & Visual/UX Designer & 38 (3.4\%) \\
 & Infrastructure & 14 (1.3\%) \\
 & Marketeer or sales & 2 (0.2\%) \\
 & Other & 81 (7.2\%) \\
 & Not stated & 52 (4.7\%) \\
Age Diversity & 18-25 years & 79 (7.1\%) \\
 & 26-35 years & 494 (44.2\%) \\
 & 36-45 years & 308 (27.5\%) \\
 & 46-55 years & 128 (11.4\%) \\
 & 56-65 years & 39 (3.5\%) \\
 & 66+ years & 3 (0.3\%) \\
 & Not stated & 52 (4.7\%) \\
Cultural Diversity & Western Europe & 581 (52\%) \\
 & Eastern Europe & 104 (9.3\%) \\
 & North America & 103 (9.2\%) \\
 & Central \& South America & 48 (4.3\%) \\
 & Middle East & 30 (2.7\%) \\
 & South-East Asia & 28 (2.5\%) \\
 & South Asia & 27 (2.4\%) \\
 & East Asia & 20 (1.8\%) \\
 & Oceania & 7 (0.6\%) \\
 & Africa & 2 (0.2\%) \\
 & Other & 51 (4.6\%) \\
 & Not stated & 11 (1\%) \\
Gender Diversity & 100\% men or women & 174 (15.6\%) \\
 & 80\% men and 20\% women & 766 (68.5\%) \\
 & 50\%-50\% men and women & 155 (13.9\%) \\
 & 20\% men and 80\% women & 16 (1.4\%) \\
\bottomrule
\end{tabular}
\end{table}

\subsection{Measurements}
\label{sec:measurements}

\textbf{Age, gender, role, and cultural diversity}: To assess the impact of diversity on team effectiveness, we identified three dimensions of demographic diversity that are commonly studied (age, cultural background, and gender) and one informational dimension (role). The questions and the available categories are shown in Table 1 in the Supplementary Materials. 

We measured \textit{age diversity} by asking participants to categorize themselves into an applicable age range (18-25, 26-35, 36-45, 46-55, 55-65, and 66+ years). 

For \textit{role diversity}, participants were asked to categorize their work into the most applicable software team role (e.g. developer, tester, designer, infrastructure) or a ``Other''-category

\textit{Cultural diversity} is often operationalized through ethnicity as a proxy for differences in value systems, norms, and beliefs. We chose against this for two reasons. The first was that several worker unions of participating organizations objected to such measures during a pilot. Moreover, the European General Data Protection Regulation (GDPR) considers ethnicity as ``particularly sensitive'' personal data and generally prohibits its collection~\cite{EU_Commission_Sensitive_Data}. The second reason is that ethnicity is a surface-level variable that does not imply cultural diversity. Indeed, a study by Desmet, Ortuño-Ortín \& Wacziarg~\cite{desmet2017culture} in 76 countries showed that people with different ethnicities who live in the same country are more similar in value systems than people with the same ethnicities who live in different countries. Thus, we opted for another way to assess cultural diversity. Participants were asked to identify the world region where they had lived the longest (e.g. Western Europe, North America, South-East Asia). This reflects a more dynamic understanding of cultural diversity than ethnicity or place of birth because it recognizes cultural mobility~\cite{portes2006immigrant} and cultural exposure in identity formation~\cite{schwartz2013identity}. This assumes that the region where one has lived the longest most substantially shapes the value system one brings to a team.

Because our study is aimed at team-level diversity, we aggregated individual indicators for age, role, and cultural background to a team-level Gini-Simpson coefficient. This coefficient is a statistical indicator of the diversity of the members in a sample, ranging between 0 (no diversity) and 1 (maximum diversity) \cite{simpson1949measurement}. 

\textit{Gender diversity} was measured directly at the team level. Similarly to cultural diversity, several worker councils objected to questions that measured gender at the individual level during our pilot study. We also observed that many participants left the question unanswered. To prevent this measure from becoming an obstacle to participation, we instead asked the initiating participant of each team to indicate the gender distribution at the team level (\% women and men). We recognize there are more genders. However, we had to take this shortcut to obtain a reliable statistical analysis. 

\textbf{Team Effectiveness} was operationalized similarly to Verwijs \& Russo~\cite{verwijs2023theory}. \textit{Team effectiveness} is often defined as \textit{``the degree to which a team meets the expectations of the quality of the outcome''}~\cite{hackman1976design}. In this sense, stakeholder satisfaction is the evaluation of team outcomes from the external perspective of stakeholders (e.g., clients, customers, and users), whereas team morale is the evaluation of team outcomes from the internal perspective of team members. This is conceptually similar to how team effectiveness is defined in the ``Team Diagnostic Survey (TDS)''~\cite{wageman2005team}. For team morale, we used 3 items from the ``Utrecht Work Engagement Scale'' (UWES) scale~\cite{schaufeli2002measurement} that were modified for use in teams by Van Boxmeer et al.~\cite{van2007direct}. For stakeholder satisfaction, we used a 4-item scale developed by the authors for another study~\cite{verwijs2023theory}. Both measures are self-reported. Reliability analysis (Cronbach's alpha) showed that Team Morale ($\alpha=.910$) and Stakeholder Satisfaction ($\alpha=0.832$) were consistently measured across participants.

\textbf{Relational conflict} was operationalized by adapting three items from a scale developed by Jehn et al. ~\cite{jehn1995multimethod} to measure relationship conflict. Such conflicts represent interpersonal incompatibilities between team members that ``typically includes tension,
animosity, and annoyance among members within a group''~\cite[p. 258]{jehn1995multimethod}. The items were adapted for use in teams by the authors. The reliability of measurements across participants was high ($\alpha=.892$). 

\textbf{Psychological Safety} was operationalized by adapting three items from the ``Inquiry \& Dialogue'' scale that was developed by Marsick \& Watkins~\cite{marsick2003demonstrating} as part of the \textit{Dimensions of Organizational Learning Questionnaire (DLOQ)}. The items were adapted for use in teams by the authors. The reliability of measurements across participants was high ($\alpha=.791$).

\textbf{Control Variables}: We included two items from the social responsibility scale (SDRS5)~\cite{hays1989five} to control for socially desirable answers and to control for common method bias~\cite{simmering2015marker}. Four categorical items were included to control for contextual variables that might influence both our independent and our dependent variables; team size, organization size, product type, and organization sector. Their categories are shown in Table~\ref{tab:samplecomposition}.

The measurement reliability of our scales is summarized in Table~\ref{tab:scales}.

\subsection{Analysis}
\label{sec:analysis}
We employed Structural Equation Modeling (SEM) with the AMOS software package~\cite{arbuckle2011ibm} to analyze the data. A strength of SEM is that it is an inherently confirmatory approach that combines multiple linear regressions and confirmatory factor analysis (CFA) with Maximum Likelihood estimation (ML) to produce more consistent and less biased estimates than those derived through Ordinary Least Squares (OLS) that is typically used in multiple regression and ANOVA~\cite{hair2019multivariate}. Furthermore, SEM allows researchers to simultaneously test both the structural part of a theory - the relationships between independent and dependent variables - and the measurement model - the inclusion of multiple indicators to measure latent factors~\cite{hair2019multivariate,byrne2010structural,russo2021PLS}. This is particularly useful for psychometric scales that use multiple questions to operationalize an underlying construct, as we do in this study.

In SEM, the statistical model is evaluated through several ``Goodness of Fit'' indices and the statistical significance and effect size of individual paths. The aim is to arrive at a model as parsimonious as possible while providing a good fit (and thus explanatory power). We discuss the fit indices in section~\ref{sec:modelfitevaluations}

Next, we tested our data for the necessary statistical assumptions required for Structural Equation Modeling. First, we assessed normality by comparing our independent, dependent, and control variables against recommended thresholds for kurtosis ($<3$) and skew ($<2$)~\cite{de2014applications} in literature. This was satisfactory for all variables except cultural diversity, whose distribution was strongly leptokurtic. This means that only a few teams showed some heterogeneity in cultural diversity, whereas most were completely homogeneous. Although statistical transformations can re-normalize such distributions, this also inevitably complicates their interpretation, especially the comparison with other effects in a model~\cite{hair2019multivariate}. 

Our measure for gender diversity was not continuous but ordinal (no diversity, some diversity, or high diversity). We treated this variable as continuous in our analyses because such a model is more parsimonious than one that treats it as categorical \cite{long2006regression}. It also simplifies the interpretation and retains more information than a model where the ordinal variable is treated as categorical. However, this requires that the relationship between the dependent variable and the ordinal independent variable is linear and that each step is approximately evenly spaced \cite{long2006regression}. The relationship was linear, but the steps were relatively not evenly spaced (respectively $.33$ and $.18$ for both steps). However, the modest violation did not warrant using a less parsimonious model with categorical dummy variables for gender diversity instead of a single variable.

We assessed homoscedasticity by inspecting the scatter plots for all pairs of independent and dependent variables for inconsistent patterns but found none. %Finally, we assessed the assumption of homoscedasticity by analyzing the residual plots for our dependent variables. The plots for our diversity indicators and our control variable for experience with Scrum indicated linear relationships. However, the residuals for our control variable for psychological safety showed modest heteroscedasticity. A consequence of this is reduced statistical power due to underestimation of the standard errors \cite{hair2019multivariate}.
Finally, multicollinearity was assessed by entering all independent variables one by one into a linear regression~\cite{gaskin2012data}. The Variance Inflation Factor (VIF) remained below the critical threshold of 10~\cite{hair2019multivariate} for all measures, ranging between 1.01 and 2.75.

Using a single method - like a questionnaire - introduces the potential for a systematic response bias where the method itself influences answers~\cite{podsakoff2003common}. To control for such common method bias, the recommended approach in current literature is using a marker variable that is theoretically unrelated to other factors in the model~\cite{simmering2015marker}. We included two items from the social responsibility scale (SDRS5)~\cite{hays1989five} and found a small but significant unevenly distributed response bias. Following recommendations in the literature, we retained the marker variable ``social desirability'' in our causal model to control for common method bias~\cite{simmering2015marker}.

We created a full latent variable model containing both the measurement and structural models. The measurement model defines relationships between indicator variables (survey items) and underlying first-order latent factors and effectively acts as a CFA-model~\cite{kline2015principles}. The structural model defines the hypothesized relations between latent variables and is a regression model. This approach makes the results less prone to convergence issues because of low indicator reliability and offers more degrees of freedom to the analysis compared to a non-latent model~\cite{harring2012comparison}. We began by assessing the measurement model following the approach outlined in literature~\cite{byrne2010structural,russo2021PLS,hair2019multivariate}. Psychological safety, relational conflict, team morale, and stakeholder satisfaction were entered as first-order latent factors, with their respective survey items as indicator variables. Once the measurement model exhibited a good fit (see section~\ref{sec:modelfitevaluations}), we added the structural part of the model.

In the structural part of the model, we created a second-order latent factor to reflect the composite nature of ``team effectiveness''. The first-order latent factors for team morale and stakeholder satisfaction were modeled as indicators, similar to Verwijs \& Russo~\cite{verwijs2023theory}. We calculated interaction terms by multiplying each team's standardized factor score for psychological safety with their standardized scores for each diversity indicator (age, gender, role, cultural background)~\cite{hayes2017introduction}. The diversity indicators and the interaction terms were entered into the model as exogenous variables. The exogenous variables for psychological safety, the diversity indicators, and their interaction terms were allowed to co-vary. No covariances were allowed between endogenous variables as our model predicted specific paths between them.

\begin{table*}[!ht]
\small
\centering
\caption{Scales used in the survey study, along with attribution, number of items, and reliability (Cronbach's Alpha) based on respondent-level response data ($N=1,118$)}
\label{tab:scales}
\begin{tabularx}{\textwidth}{@{}lXll@{}}
\toprule
\textbf{Construct variable} & \textbf{Items adapted from} & \textbf{\# Items} & \textbf{Alpha} \\ 
\midrule
Psychological Safety & Adapted from ``Inquiry \& Dialogue'' scale in DLOQ~\cite{marsick2003demonstrating} & 5 & .791 \\
Relational Conflict & Adapted from Jehn~\cite{jehn1995multimethod} & 3 & .892 \\
Stakeholder Satisfaction & Created by authors from our case studies & 4 & .874 \\
Team Morale & Adapted from Van Boxmeer et. al.~\cite{van2007direct} and Schaufeli~\cite{schaufeli2002measurement} & 3 & .910 \\
Social Desirability & Highest-loading items from SDRS-5 scale~\cite{hays1989five} & 2 & .672 \\ 
\bottomrule
\end{tabularx}
\end{table*}

\subsection{Model fit evaluations}
\label{sec:modelfitevaluations}
We assessed reliability, convergent, and discriminant validity for the resulting measurement model before testing for the model fit. The individual steps involved in the model-fitting process are in Table 2 in the Appendix. Discriminant validity was assessed by analyzing the heterotrait-monotrait ratio of correlations (HTMT) with a third-party plugin in AMOS~\cite{gaskin2016master} and following the approach outlined in literature~\cite{hair2019multivariate,henseler2015new}. This ratio between trait correlations and within trait correlations should remain below $R=.90$ to indicate good discriminant validity from other constructs in different settings. This was the case for all measures.
We assessed convergent validity by inspecting composite reliability (CR) and average extracted variance (AVE). The AVE remained above the rule of thumb of $>.50$~\cite{hair2019multivariate} for all pairs of factors, ranging between .621 and .890. The CR was equal to or above the threshold of .70~\cite {hair2019multivariate} for all scales.

We then proceeded with the fitting procedure. We investigated local fit by inspecting the residual covariance matrix. A standardized residual covariance is considered large when it exceeds 2.58~\cite{byrne2010structural}. This indicates that an item does not sufficiently measure (only) its intended factor. One item from Stakeholder Satisfaction (StakeholderSatisfaction3) showed poor local fit, and we removed it. 

The overall goodness of fit was evaluated with indices recommended by recent literature ~\cite{kline2015principles,byrne2010structural,hair2019multivariate}; the Comparative Fit Index (CFI), the Root Mean Error of Approximation (RMSEA), the Standardized Root Mean Residual (SRMR) and the Tucker Lewis Index (TLI). The Comparative Fit Index (CFI)~\cite{bentler1990comparative} offers a similar test to CMIN/df but with consideration of the sample size and its reliable properties have made it the most commonly used index today~\cite{hair2019multivariate}. A cut-off value of .95 or higher is generally considered to indicate good fit~\cite{byrne2010structural,hair2019multivariate,bentler1980multivariate}. The Root Mean Error of Approximation (RMSEA) by Steiger \& Lind~\cite{steiger1980statistically} also provides an index that considers sample size but adds to this a parsimony adjustment that leads it to favor the simplest model out of potential models with the same explanatory power~\cite{kline2015principles}. A value below .05 is generally considered to indicate a good fit~\cite{byrne2010structural, hair2019multivariate}; additionally, we follow the advice to report the confidence interval in addition to only the absolute value~\cite{chen2008empirical}. The Standardized Root Mean Residual (SRMR) calculates a standardized mean of all the differences (residuals) between each observed covariance and the hypothesized covariance between variables~\cite{hair2019multivariate}. A value below .08 is indicative of a good fit. We also inspected local fit by looking at the standardized residuals between pairs of variables, with values beyond 2.58 as a cut-off value for poor local fit~\cite{byrne2010structural}. Finally, we report and test the Tucker-Lewis Index (TLI). This is another incremental fit index, like the CFI, that compares the relative improvement of the hypothesized model from a model where all variables are uncorrelated. Hair et al. ~\cite{hair2019multivariate} considers a value of .97 or above sufficient to conclude a good model fit. In addition to overall model fit, we also evaluated our model on the percentage of variance that is explained in team effectiveness by all other variables in the model. 

The measurement model fitted our data well ($Chi^2(79)=127.650$; $TLI=.973$; $CFI=.980$; $RMSEA=.062$; $SRMR=.0516$). A Confirmatory Factor Analysis (CFA) is reported in the Appendix (Table 3) that shows that all items loaded primarily on their intended factors, except for the item PsychologicalSafety1. This item also loaded negatively on the factor for Relational Conflict. The cumulative Eigenvalues of 5 factors explain 78\% of the total observed variance, which is well beyond the recommended threshold of 60\%~\cite{hair2019multivariate}. 

We then tested the path model for the effects we predicted from our theory. 
Our hypothesized theoretical model fits the data well on each fit indices, as described in Table~\ref{tab:modelfitindices}: $Chi^2(129)=156.282$; $TLI=.981$; $CFI=.988$; $RMSEA=.036$; $SRMR=.051$. The predictors in our model explain respectively 40.7\% of the variance in the latent factor representing team effectiveness. For studies in the social sciences, values above 26\% are considered large~\cite{cohen1992power}. All steps of the fitting process are listed in Table 2 of the Appendix.

%Finally, an alternative model was tested that controlled for team size, organization size, organization sector, and product type. When the inclusion of control variables does not meaningfully change path coefficients or their significance, we can more robustly attribute changes in dependent variables to our independent variables and not the variation in control variables~\cite{kline2015principles,byrne2010structural}. The control variables were entered into the path model and loaded onto all indicators. This alternative model fitted the data less well than our primary model ($Chi^2(155)=210.603$; $TLI=.954$; $CFI=.976$; $RMSEA=.047$; $SRMR=.0608$). The changes in path coefficients were minor, ranging between $0.01$ and $0.08$, and no changes were observed in path significance. Of the control variables, product type and organization size did not significantly influence any variable in our model. A small effect was observed from team size ($\beta=.160,p<.05$) and organization sector ($\beta=.005,p<.01$) on gender diversity, but not on the dependent variables. Thus we conclude that variations in these control variables do not provide alternative explanations for our findings. 

To enhance the generalizability of our model, we extended our analysis to include four control variables: team size, organization size, sector, and product type. The aim of this test was to verify that the inclusion of these control variables did not meaningfully alter the path coefficients or their significance~\cite{kline2015principles, byrne2010structural}. The resulting alternative model, which integrated these control variables, demonstrated lower fit compared to the original model ($Chi^2(155)=210.603$; $TLI=.954$; $CFI=.976$; $RMSEA=.047$; $SRMR=.0608$). This means that the uncontrolled model is more parsimonious with similar or better predictive power. Crucially, the inclusion of these variables did not alter the relationships within our model in a meaningful way. This suggests that our findings are robust and generalizable, and unaffected by the team-level and organization-level factors we controlled for.

\begin{table*}[]
\centering
\caption{Model Fit Indices}
\label{tab:modelfitindices}
\begin{tabular}{@{}m{8cm}m{2cm}m{8cm}@{}}
\toprule
\textbf{Model fit index}                                & \textbf{Value}      & \textbf{Interpretation}                                                                                                                                             \\ \midrule
Chi-Square ($\chi^2$)                                            & 156.282            & n/a                                                                                                                                                                 \\ 
Degrees of freedom (df)                                 & 129                 & n/a                                                                                                                                                                 \\ 
CMIN/df                                                 & 1.211               & A value below 5 indicates an acceptable model fit~\cite{marsh1985application}, below 3 a good fit~\cite{hu1999cutoff}
\\ 
Root Mean Square Error of Approximation (RMSEA) & .036 & Values $\leq .05$ indicates good model fit~\cite{byrne2010structural}  \\                                RMSEA 90\% Confidence Interval   & .000-.055                                                                            \\ 
$p$ of Close Fit (PCLOSE)                                 & .873               & Probability that RMSEA $\leq 0.05$, where higher is better                                                                                                    \\ 
Comparative Fit Index (CFI)                             & .988               & Values $\geq .97$ indicates good model fit~\cite{hair2019multivariate}                                                                                                               \\ 
Tucker Lewis Index (TLI)                                & .981               & Values $\geq .97$ indicates good model fit~\cite{hair2019multivariate}                                                                                                                \\ 
Standardized Root Mean Square Residual (SRMR)           & .051              & Values $ \leq .08$ indicates good model fit~\cite{hair2019multivariate}                                                                                                                    \\ 
Variance explained by predictors ($R^2$) of Team Effectiveness          & 40.7\%              & Values $ \geq 26\%$ indicates large effect~\cite{cohen1992power}                                          
\\ \bottomrule
\end{tabular}
\end{table*}

\begin{table*}[]
\tiny
\centering
\caption{Means, Standard Deviations, Skewness, Kurtosis and Correlations (Pearson) for continuous variables. Correlations marked with * are significant at $p < 0.05$}.
\label{tab:results}
\resizebox{\textwidth}{!}{%
\begin{tabular}{@{}Xllllllllllllll@{}}
\toprule
 & \textbf{Variable} & \textbf{Mean} & \textbf{SD} & \textbf{Skewness} & \textbf{Kurtosis} & \textbf{1} & \textbf{2} & \textbf{3} & \textbf{4} & \textbf{5} & \textbf{6} & \textbf{7} & \textbf{8} \\ \midrule
1 & Gender Diversity & 1.97 & .55 & -.02 & .33 & 1.00 &  &  &  &  &  &  &   \\
2 & Age Diversity & .71 & .20 & -.78 & .75 & .26* & 1.00 &  &  &  &  &  &   \\
3 & Cultural Diversity & .05 & .16 & 2.97 & 7.98 & .03 & .00 & 1.00 &  &  &  &  &  \\
4 & Role Diversity & .61 & .22 & -1.06 & 1.01 & -.07 & .02 & -.06 & 1.00 &  &  &  &  \\
5 & Psychological Safety & 5.54 & .66 & -.80 & 1.29 & .10 & -.06 & .04 & -.03 & 1.00 &  &  &  \\
6 & Team Effectiveness & 5.36 & .71 & -.38 & -.50 & .17 & .06 & .08 & .04 & .72* & 1.00 &  &  \\
7 & Relational Conflict & 2.45 & .98 & 1.16 & 1.56 & .06 & .00 & -.01 & -.07 & -.71* & -.50* & 1.00 &  \\
\multicolumn{14}{c}{\emph{Control Variables}} \\ 
8 & Social Desirability & 5.72 & .48 & -.17 & -.03 & .16 & .11 & .02 & -.02 & .59* & .56* & -.40* & 1.00 \\
\bottomrule
\end{tabular}
}
\end{table*}

\section{Results}
\begin{figure*}
\centering

\includegraphics[height=3in]{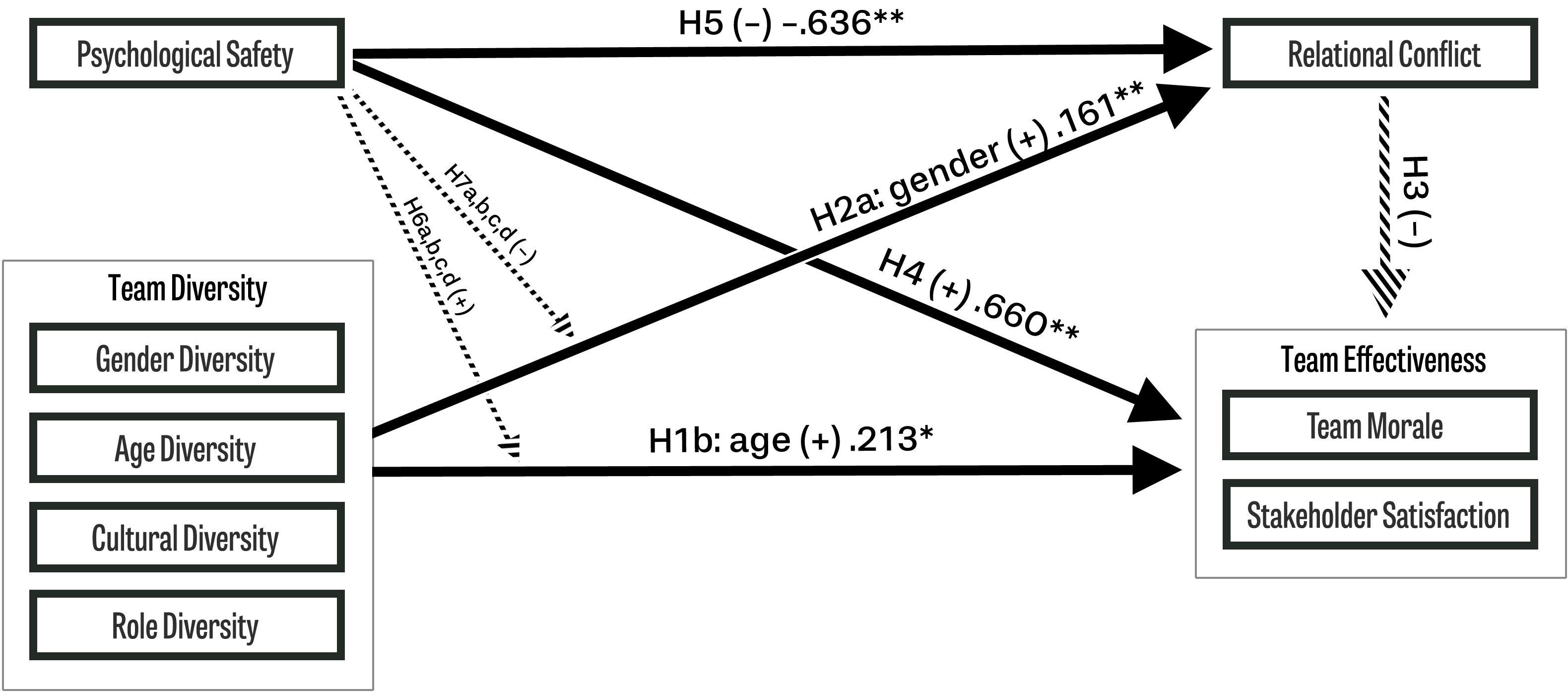}
\caption{Standardized path coefficients for the model ($**: p < .01$, $*: p < .05$). 
%The Squared Multiple Correlation for team effectiveness is reported at its top right. 
The dotted lines represent non-significant results. Indicator items and non-significant paths for sub-hypotheses are omitted to improve readability. A detailed overview of the individual hypotheses is reported in Table~\ref{tab:parameterestimates}.}
\label{fig:results}
\end{figure*}

We now turn to the results and hypothesis testing. The means, standard deviations, and Pearson correlations of all variables are reported in Table~\ref{tab:results}. Significant effects are also visualized in Figure~\ref{fig:results}.
Following recommendations in statistical literature~\cite{byrne2010structural,kline2015principles}, we used a bootstrapping procedure with 2,000 samples and 95\% bias-corrected confidence intervals to more accurately estimate parameters and their \textit{p}-values for direct effects, factor loadings, and the hypothesized indirect effects. This resulted in a standardized, bias-corrected estimate ($\beta$) for each path, along with a \textit{p}-value to test whether the null hypothesis can be rejected. $\beta$  represents the change in the predicted variable in standard deviation units for a one standard deviation change in the predictor variable while holding all other variables in the model constant. They can be used to compare the strength of an effect as compared to other effects in the same model~\cite{byrne2010structural}.

The parameter estimates relevant to our hypotheses are reported in Table~\ref{tab:parameterestimates}.

\begin{table*}[!ht]
\centering
\small
\caption{Parameter Estimates, Confidence Intervals, Standard Errors, Standardized Coefficients for Direct Effects, Interaction Terms and Indirect effects for hypotheses, and Factor Loadings. Significant effects are marked with $**: p < .01$, $*: p < 0.05$}
\label{tab:parameterestimates}
\begin{tabularx}{\textwidth}{@{}Xlllll@{}}
\toprule
\textbf{Parameter} & \textbf{Unstandardized} & \textbf{95\% CI} & \textbf{SE} & \textit{\textbf{p}} & \textbf{Standardized} \\ \midrule
\multicolumn{6}{c}{\emph{Direct Effects}} \\ \addlinespace   
H1a: Gender Diversity $\rightarrow$ Team Effectiveness & .037 & (-.075, .123) & -.075 & .622 & .056 \\
H1b: Age Diversity $\rightarrow$ Team Effectiveness* & .391 & (.077, .800) & .077 & .041 & .213 \\
H1c: Cultural Diversity $\rightarrow$ Team Effectiveness & .024 & (-.382, .542) & -.382 & .872 & .010 \\
H1d: Role Diversity $\rightarrow$ Team Effectiveness & .022 & (-.315, .290) & -.315 & .956 & .013 \\
H2a: Gender Diversity $\rightarrow$ Relational Conflict** & .241 & (.108, .417) & .108 & .008 & .161 \\
H2b: Age Diversity $\rightarrow$ Relational Conflict & -.490 & (-1.178, .026) & -1.178 & .118 & -.117 \\
H2c: Cultural Diversity $\rightarrow$ Relational Conflict & .032 & (-.824, 1.022) & -.824 & .855 & .006 \\
H2d: Role Diversity $\rightarrow$ Relational Conflict & -.332 & (-.834, .199) & -.834 & .306 & -.087 \\
H3: Relational Conflict $\rightarrow$ Team Effectiveness & .035 & (-.091, .181) & -.091 & .747 & .081 \\
H4: Psychological Safety $\rightarrow$ Team Effectiveness** & .574 & (.300, .927) & .300 & .004 & .660 \\
H5: Psychological Safety $\rightarrow$ Relational Conflict** & -1.262 & (-1.888, -.727) & -1.888 & .001 & -.636 \\ \addlinespace 
\multicolumn{6}{c}{\emph{Interactions}} \\ \addlinespace   
H6a: Gender Diversity * Psychological Safety $\rightarrow$ Team Effectiveness & -.018 & (-.093, .037) & -.093 & .550 & -.052 \\
H6b: Age Diversity * Psychological Safety $\rightarrow$ Team Effectiveness & .007 & (-.069, .086) & -.069 & .885 & .020 \\
H6c: Cultural Diversity * Psychological Safety $\rightarrow$ Team Effectiveness & -.042 & (-.159, .061) & -.159 & .388 & -.081 \\
H6d: Role Diversity * Psychological Safety $\rightarrow$ Team Effectiveness & .026 & (-.070, .082) & -.070 & .550 & .076 \\
H7a: Gender Diversity * Psychological Safety $\rightarrow$ Relational Conflict & -.057 & (-.138, .054) & -.138 & .416 & -.072 \\
H7b: Age Diversity * Psychological Safety $\rightarrow$ Relational Conflict & .014 & (-.095, .129) & -.095 & .812 & .017 \\
H7c: Cultural Diversity * Psychological Safety $\rightarrow$ Relational Conflict & .133 & (-.047, .377) & -.047 & .206 & .112 \\
H7d: Role Diversity * Psychological Safety $\rightarrow$ Relational Conflict & -.044 & (-.162, .046) & -.162 & .406 & -.057 \\ \addlinespace 
\multicolumn{6}{c}{\emph{Factor loadings from first to second-order factors}} \\ \addlinespace 
Team Effectiveness $\rightarrow$ Stakeholder Happiness** & .752 & (.144, .614) & .114 & .003 & .389 \\
Team Effectiveness $\rightarrow$ Team Morale & 1.000 & (.618, 1.800) & & & .873 \\ 
\bottomrule
\end{tabularx}
\end{table*}

Our results allowed us to reject the null hypotheses for 4 out of 19 (sub)hypotheses. The primary hypothesis of this study is that diversity makes software teams more effective because it broadens the cognitive resources available for information processing (H1a-d). This is partially true for our results, as only age diversity significantly contributes to team effectiveness (H1b, $\beta=.213,p<.05$). Software teams seem slightly more effective when there is greater age heterogeneity. In contrast, heterogeneity in gender, cultural background, or role does not appear relevant to team effectiveness (H1a,c,d).

We also hypothesized that relational conflict in teams would increase as heterogeneity increases and members become less similar. This is also partially true, as only gender diversity significantly contributes to relational conflict (H2a, $\beta=.161,p<.01$). Thus, there appears to be more conflict as teams grow more heterogeneous in gender. There is no discernible effect of diversity in age, cultural background, or functional role on conflict. (H2b,c,d).

Contrary to our expectations, we did not find a significant effect between relational conflict and team effectiveness (H3). Teams that experience more relational conflict do not seem to be more or less effective than teams that experience less conflict. However, the results show a strong positive effect of psychological safety on team effectiveness (H4, $\beta=.660,p<.01$). Teams that experience more psychological safety are more effective in that they have reported more satisfied stakeholders and higher team morale. Psychological safety also strongly decreases the amount of relational conflict reported by teams (H5, $\beta=-.636,p<.01$).

Finally, we hypothesized that psychological safety moderates the strength by which diversity contributes to team effectiveness and relational conflict. However, none of the interactions were significant (H6a-d; H7a-d).

\section{Discussion}
\label{sec:discussion}
This study investigated how diversity in age, role, cultural background, and gender influences the effectiveness of software teams. 1,118 respondents from 161 software teams participated in our study. Overall, our results provide mixed support for both the benefits and the risks of member heterogeneity in teams. A summary of our findings is provided in Table~\ref{tab:Findings}.

According to the \textit{categorization-elaboration model (CEM)}\cite{van2004work} and \textit{cognitive resource diversity theory}, we hypothesized that software teams benefit from diversity as it expands the cognitive resources available for information processing. However, only age diversity improves team effectiveness directly. In other words, teams are more effective when their members vary in age. This is probably a proxy for differences in tenure and experience that encourage innovation and creativity~\cite{pesch2015effects}. However, generational differences in work values have also been found to be relevant~\cite{wey2002generational}. Either way, this is in support of \textit{cognitive resource diversity theory} and its prediction that diversity expands cognitive resources that teams have access to.

Our findings are consistent with the conclusions from a recent review of the literature by Tshetshema \& Chan~\cite{tshetshema2020systematic}, and a meta-analysis of 74 studies by Schneid et al. ~\cite{schneid2016age}, particularly for complex tasks. However, another meta-analysis of 35 studies by Horwitz \& Horwitz~\cite{horwitz2007effects} found no positive impact of demographic diversity (age, gender, race). So our results are more nuanced than the overall positive effect of team diversity that is reported by Lee \& Xia~\cite{lee2010toward} for software teams. We also did not find a positive effect of gender diversity or cultural diversity, whereas others did~\cite{tshetshema2020systematic,russo2020gender}. All in all, the association between demographic diversity and team effectiveness is more complicated than the direct, positive effects we hypothesized.

In addition to demographic diversity, we also investigated how role diversity improves team effectiveness. Agile software methodologies in particular emphasize this type of diversity as an important characteristic of autonomous teams~\cite{schwaber2020scrum,larman2004agile}. In line with \textit{cognitive resource diversity theory}, role diversity allows teams to leverage more perspectives and broader informational resources to resolve complex problems~\cite{van2004work,horwitz2005compositional}. When members bring more functional roles to their work together (e.g., analyst, tester, developer, designer), their shared mental models will be richer than when all members hold the same role (e.g., developer)~\cite{bowers2000member,cannonbowerssalas1993}. However, we did not find evidence for this. Teams with high role diversity were not more or less effective than teams with lower role diversity. This is partially consistent with extant literature. Homberg \& Bui~\cite{homberg2013top} found no evidence for a link between role diversity and team effectiveness in a meta-analysis of other empirical studies. Horwitz \& Horwitz~\cite{horwitz2007effects} also did not find an effect on team performance, although they did find one on the quality of work done by teams.

Diversity in teams is often considered a double-edged sword in the literature on diversity~\cite{horwitz2005compositional}. The CEM proposes that diversity can also harm team effectiveness through the \textit{similarity-attraction paradigm}~\cite{williams1998demography}. As members grow less similar and bring different perspectives to teamwork, there is more potential for tension and conflict. This decreases the ability of teams to elaborate information effectively and reduces their effectiveness. Concerning the first assertion, our results show that gender diversity does increase relational conflict but not other kinds of diversity. This finding is consistent with some studies~\cite{haas2010can}, but not others~\cite{russo2020gender,tshetshema2020systematic}. Regarding the second assertion, we failed to find any impact of relational conflict on team effectiveness. So while it appears true that gender diversity increases relational conflict in teams to some extent, we cannot conclude that this also harms team effectiveness (i.e., the double-edged sword).

The CEM attempts to reconcile the conflicting results by drawing attention to social- and task-related moderators that shape how diversity impacts team effectiveness. We investigated one social moderator frequently associated with diversity, relational conflict, and team effectiveness: psychological safety. We hypothesized that a psychologically safe environment would make it easier for diverse teams to elaborate on task information effectively. Although psychological safety reduced relational conflict and improved team effectiveness, we could not reject the null hypotheses for psychological safety as a moderator of the diversity-effectiveness link.
%So how do our findings match with the CEM? This study is not intended as a comprehensive test of the \textit{categorization-elaboration model}. We did not directly measure processes at the heart of the CEM, like information elaboration and social categorization. However, we extrapolated core hypotheses from the CEM and tested them with suitable operationalizations. 
In summary, our results show some benefits of diversity (age) on team effectiveness and some risks of diversity through relational conflict (gender). Psychological safety also reduces relational conflict and increases team effectiveness, but we found no evidence for a moderating role in the diversity-effectiveness link or the diversity-conflict link.

\subsection{Alternative explanations}
The mixed evidence suggests that there are factors at work that moderate or mediate the effects of diversity on effectiveness and conflict. Diversity alone does not make teams more effective because it broadens cognitive resources, just as it does not inherently and consistently create conflict because members are less similar. 

This study investigated psychological safety as one potential social moderator of the diversity-effectiveness link. Our mixed results suggest that other moderators are at play. One example of this is task interdependence. A core element of Agile software methodologies is that teams work together on complex tasks~\cite{schwaber1995,larman2004agile,manifesto2001agile}. Collective elaboration of task-related information and the pooling of skills to accomplish tasks is also a common thread in the definition of teamwork~\cite{hackman1976design,schmutz2019effective}. Without task interdependence, the two mechanisms of diversity diminish. Because there is less collective elaboration, the benefits of the broadened cognitive resources that are offered by diversity diminish. Furthermore, a major source of conflict between members is removed because they spend much less time together processing information. Members may have more ``skin in the game'' when they feel they depend on others in their team to be successful. Paradoxically, this may surface as a higher degree of relational conflict than teams with very low interdependence. In this sense, psychological safety is likely only relevant as a moderator of the diversity-effectiveness link in teams with high task interdependence but not low task interdependence. Future studies can investigate if the effects of diversity and psychological safety are indeed more pronounced when controlling for task interdependence.

Another explanation may be that the effect of diversity on team effectiveness is not linear. Several authors~\cite{earley2000creating,jetten1998defining} have argued for curvilinear models where diversity contributes to effectiveness only when it is moderated (inverted U) or when it is either low or high (upright U). Which model applies varies by diversity type. For example, Dahlin, Weingart \& Hinds~\cite{dahlin2005team} found that educational diversity contributed to team performance when it was either low or high (upright U) but found the opposite for national diversity (inverted U). Richard et al. ~\cite{richard2004cultural} found that management teams with moderate gender diversity performed better than teams with low or high diversity, but only in high-risk settings (inverted U). However, diversity in terms of age, gender, or function may contribute to learning behavior in teams more strongly when diversity is low or high but not moderate (upright U)~\cite{gibson2003healthy}. So while there is some support for the curvilinear effects of diversity, the relationship is complex. To further complicate matters, the shape of the relationship may also be moderated by the expectations that teams themselves have of the benefits of diversity~\cite{van2007work}. We performed a posthoc test to assess whether a curvilinear relationship between dimensions of diversity and team effectiveness better fitted the data. This was not the case. A quadratic regression model was not significant for the following diversity dimensions: age ($R^2 = .004, F(2, 158) = .321, p =  .726$), gender ($R^2 = .021, F(2, 158) = 1.695, p = .187$), culture ($R^2 = .008, F(2, 158) = .664, p = .516$), and role ($R^2 = .000, F(2, 158) = .025, p = .975$). Thus, the possibility of a curvilinear relationship rather than a linear one does not appear to explain the lack of results in this study.

We often assume that diversity in age, gender, function, and cultural background inherently leads to a different understanding of the task and potential solutions. This is both the strength and the weakness of diverse teams. In the day-to-day practice of teams, such differences in understanding may also lead to conflict if members need to adequately express their views and integrate them with other members into a synthesized solution. In addition to the task-related and social moderators mentioned above, it is reasonable to expect that communication and conflict navigation skills are also highly relevant, as well as the presence of an environment where such different understandings can be elaborated effectively. Few studies have investigated such moderators, particularly for software teams~\cite{silveira2019systematic}. Furthermore, this ties into team members' beliefs about diversity, how to deal with it, and whether or not it benefits teamwork. Van Knippenberg et al. ~\cite{van2005diversity,van2007work} call this a ``Diversity Mind-Set''. Several studies have shown that teams and organizations can better leverage diversity when they recognize it as a strength and have learned how to appreciate and deal with the resulting informational diversity~\cite{ely2001cultural,richard2004cultural,van2007work}. 

\begin{table*}[!ht]
\centering
\caption{Summary of key findings \& implications for practice}
\label{tab:Findings}
\begin{tabular}{@{}m{3cm}m{6.8cm}m{6.8cm}@{}}
\toprule
    & \textbf{Findings} & \textbf{Implications for practice} \\
    \midrule
  \textbf{Diversity \& team effectiveness} & Based on existing theory, we developed a Structural Equation Model for how diversity and psychological safety interact to impact team effectiveness and relational conflict. The model fitted the data well ($Chi^2(129)=156.282$; $TLI=.981$; $CFI=.988$; $RMSEA=.036$; $SRMR=.051$). Age diversity showed a positive association with team effectiveness ($\beta=.213,p<.05$), but not diversity in gender, role, or cultural background. & Teams with members of different age groups will likely benefit from the broader range of tenure and work/life experience. The benefits of other types of diversity appear more conditional on moderating factors. Organizations can assess the extent to which teams are diverse. However, psychological safety, communication skills, and a diversity mindset seem important moderators that organizations need to provide and encourage teams to leverage it. \\  \addlinespace
  \textbf{Diversity \& relational conflict} & Gender diversity was positively associated with relational conflict in software teams ($\beta=.161,p<.01$). However, diversity in role, age, or cultural background did not. In turn, relational conflict did not significantly affect team effectiveness. & When teams grow more diverse, members' different perspectives may lead to more conflict and friction. This appears particularly relevant to gender diversity. Such negative consequences of diversity may be counteracted when teams learn to see their diversity as a strength and recognize that different perspectives can be reconciled through open dialogue and elaboration. \\  \addlinespace
  \textbf{Psychological safety \& team effectiveness} & Psychological safety was positively associated with team effectiveness ($\beta=.660,p<.01$) and negatively associated with relational conflict ($\beta=-.636,p<.01$) & Teams that operate in environments where members can openly and safely elaborate information are more effective than other teams, regardless of their diversity. They also experience much less relational conflict. Organizations do well to develop the skills, support structures, and management styles that foster psychological safety in and around teams \\  \addlinespace
  \textbf{Psychological safety as a moderator} & Psychological safety did not significantly moderate the association between diversity and team effectiveness, nor between diversity and relational conflict. & Psychological safety is paramount, but it does not appear to strengthen the cognitive benefits of team diversity, nor does not it appear to buffer against negative consequences. \\
    \bottomrule
\end{tabular}
\vspace{2em}
\end{table*}

\textbf{For practitioners}, it is important to notice that our results are broadly consistent with existing research, showing that team diversity is not unequivocally beneficial or harmful. Although we found a positive effect of age diversity, the effects of other types of diversity appear to be more conditional on moderating factors. Several factors have been proposed to date, like the autonomy that teams have~\cite{lee2010toward}, task difficulty~\cite{bowers2000member}, psychological safety~\cite{edmondson1999psychological}, team climate~\cite{bodla2018diversity} and the beliefs that teams have about diversity~\cite{van2007work}. \textbf{\textit{This suggests that context is just as important as diversity alone}}. The practical implications of our findings are summarized in Table~\ref{tab:Findings}.

\subsection{Limitations}
In the following section, we discuss the threats to the validity of our sample study.

\textbf{Internal validity}
Internal validity refers to the confidence with which changes in the dependent variables can be attributed to the independent variables and not other uncontrolled factors~\cite{cook1979quasi}. We employed several strategies to maximize internal validity. First, we recognized that online questionnaires are prone to bias and self-selection as a result of their voluntary (non-probabilistic) nature. We counteracted this by embedding our questions in a tool that is regularly used by software teams to self-diagnose their process and identify improvements. Team members were invited by people in their organization to participate. Second, we thoroughly cleaned the dataset of careless responses to prevent them from influencing the results. Third, we did not inform the participants of our specific research questions to prevent them from answering in a socially desirable manner. We also controlled for social desirability in participants' responses, as well as common method bias introduced when a single method is used to collect data. Finally, we tested an alternative model to rule out that variation in team size, organization sector, organization size, or product type instead explained our findings, which was not the case.

Despite our safeguards, there may still be confounding variables that we were unable to control for. This is particularly relevant to the operationalization of team effectiveness, which is based on self-reported scores on team morale and the perceived satisfaction of stakeholders. Mathieu et al. ~\cite{mathieu2008team} recognize that such affect-based measures may suffer from a ``halo effect''. Future studies could ask stakeholders to rate their satisfaction with team outcomes directly. This does not entirely rule out a halo effect but is conceptually closer to what matters to organizations. Future studies could also find more objective measures for team effectiveness.

\textbf{Construct validity}
Construct validity refers to the degree to which the measures used in a study measure their intended constructs~\cite{cook1979quasi}. We adapted items from established scales to measure psychological safety~\cite{marsick1999facilitating}, team effectiveness~\cite{verwijs2023theory}, relational conflict~\cite{jehn1995multimethod} and social desirability~\cite{hays1989five}. A confirmatory factor analysis (CFA) showed that all items were loaded primarily on their intended scales (see Table 3 in the Appendix). A heterotrait-monotrait (HTMT) analysis confirmed discriminant validity for all measures. The reliability for all measures exceeded the cutoff recommended in the literature ($CR>=.70$~\cite{hair2019multivariate}), except social desirability. Thus, we are confident that we reliably measured the intended constructs.

A limitation of our measure for team effectiveness is that it only addressed (self-reported)  stakeholder satisfaction and team morale. Although both are reasonable and relevant aspects of team effectiveness and are commonly used in team research~\cite{hackman1976design}, effectiveness is also a more-faceted construct~\cite{guzzo1996teams}. 

Our measure for role diversity captured the role for 88\% of participants. 7.2\% picked the ``Other'' category. While this probably reflects a long tail of more niche roles, a more sensitive measure with more than the seven roles currently provided would've increased the resolution for role diversity slightly.

While our sample consisted of software teams, we note that all practiced Agile methodologies. This was because the survey was advertised primarily on channels in the Agile community. Due to the focus on close collaboration and high task interdependence through shared goals in Agile methodologies, it is possible that the association between diversity, psychological safety, and relational conflict is different in other kinds of software teams. However, Agile methodologies have become so prevalent that most teams use them in one form or another~\cite{version1stateofagile}.

As with every sample study, the way we operationalized diversity influenced our results. This is particularly relevant for cultural diversity. We observed very little diversity in teams on this variable. Our operationalization assumed that the region where participants lived the longest most strongly influenced their value systems. However, since we used large regions (e.g., Western Europe, and Africa) the resolution of this measure may simply have been too low. Future studies that wish to use the same operationalization do well to expand the categories, perhaps even to individual countries. Another issue is that our operationalization is yet another proxy for cultural values. Other proxies such as ethnicity, race, and place of birth have been shown to be unreliable measures of actual cultural  diversity~\cite{brown2010conceptualizing}. A more direct measure of (cultural) value systems could have resulted in more robust differences. For example, House et. al.~\cite{house2004culture} identified 9 cultural dimensions based on quantitative data from 62 countries which partially overlap with earlier work by Hofstede~\cite{hofstede1984culture}. This includes dimensions that appear relevant to teamwork, such as ``Power distance'', ``Assertiveness'', ``Performance orientation'' and ``Uncertainty avoidance''. Several studies suggest that diversity on such dimensions impacts teamwork~\cite{bantz1993cultural,bochner1994power}. Thus, such traits offer a more promising operationalization of cultural diversity in future research.

Finally, we could not directly ask participants for their gender due to privacy concerns. So it was not possible to calculate a Gini index as we did for the diversity measures. The resulting measure was ordinal instead of continuous, limiting our analysis's resolution for this variable. Future studies would do well to use a more continuous measure of gender distribution.  

\textbf{Conclusion validity}
Conclusion validity assesses the extent to which the conclusions about the relationships between variables are reasonable based on the results~\cite{cozby2012methods}. We used Structural Equation Modeling to test the entire model simultaneously~\cite{kline2015principles, byrne2010structural}. The resulting model fits the data well on all fit indices recommended by statistical literature and explains a substantial amount of variance in the dependent variables. Our sample was also large enough to identify medium effects ($f=.15$) with a statistical power of 96\%.

We published team-level data and syntax files to Zenodo for reproducibility. 
%The cross-sectional nature of the data also does not allow conclusions about the causality of effects.

\textbf{External validity}
Finally, external validity concerns the extent to which the results actually represent the broader population~\cite{goodwin2016research}. First, we assess the ecological validity of our results to be high. Our questionnaire was integrated into a more general tool that Agile software teams use to improve their processes. Participants were invited by people in their organization, usually Scrum Masters. Thus, the data is more likely to reflect realistic teams than a stand-alone questionnaire or an experimental design. 

We acknowledge the potential limitations in the representativeness of our sample. Given the survey's public nature, participation was voluntary, which may have introduced a self-selection bias. It remains a possibility that the participating teams differed significantly from non-participants in terms of the variables studied or their interrelationships. To mitigate such biases, we implemented multiple bias-reducing strategies. We extensively promoted the survey across diverse online platforms frequented by team coaches, such as Agile coaches and Scrum Masters, as well as developers. We underscored the importance of anonymity, ensuring participants that their responses would remain confidential and not be shared with their respective teams or organizations. As an incentive, we offered teams a comprehensive profile, complete with actionable feedback. Our sample composition, as detailed in Table~\ref{tab:samplecomposition}, indicates participation from a diverse array of teams. These teams vary in experience, geographical location, and organizational type. The broad spectrum of scores across different measures further reinforces the diversity of our sample. The substantial sample size, combined with the aggregation of individual responses to team-level summaries, diminishes the variability arising from non-systematic individual biases.

\section{Conclusion}
\label{sec:conclusion}
A common thread in software methodologies, such as Agile, is their emphasis on teams as the primary units where complex work is performed. So it is not surprising that much research has focused on what makes such teams more effective (i.e.~\cite{verwijs2023theory,moe2010teamwork,moe2008understanding,melo2013interpretative,van2013Agile}). Although diversity is increasingly investigated in the broader literature on teams, scholarly knowledge on how it impacts software teams is still limited~\cite{silveira2019systematic,rodriguez2021perceived}. Such understanding can better equip organizations and teams to leverage diversity more effectively or learn when and how diversity is beneficial. Because what seems to be clear about diversity is that while it brings more extensive cognitive resources to teams, it can also bring more conflict as members become less similar~\cite{horwitz2005compositional}. Several models have been proposed to explain this ``double-edged sword'' of diversity, with the \textit{categorization-elaboration model (CEM)}~\cite{van2004work} as the most comprehensive one.

In this study, we explored how diversity impacts the effectiveness of software teams through the lens of the CEM theory. Our sample consisted of 1,118 team members representing 161 (Agile) software teams. Our results show that age diversity contributes to more effective teamwork but not diversity in gender, role, or cultural background. This may reflect the value of having more varied levels of experience in teams. Furthermore, the CEM also predicts a negative effect of diversity through social categorization and identity threat, which can surface through increased conflict. While our results support this effect, we only found evidence for gender diversity. Finally, the CEM predicts that task-related and social moderators influence the impact of diversity. One such moderator that is frequently studied is psychological safety~\cite{edmondson1999psychological}. While our results show that it contributes to more effective teamwork and less conflict in teams, it did not moderate the link between diversity on the one hand and effectiveness and conflict on the other. Thus, the presence of psychological safety in a team does not in itself allow teams to leverage their diversity better.
Despite the strong focus on role diversity and cross-functional teamwork in software  methodologies~\cite{schwaber1995,larman2004agile}, we found no apparent effect on team effectiveness. So while our results are broadly consistent with the CEM for age and gender diversity, it is surprising that heterogeneity in role or cultural background did not produce similar effects. One moderator that may be particularly relevant here for future research is task interdependence. 
Teams vary broadly in the degree to which members actually (need to) work together on tasks and, thus, the opportunities that arise to leverage the broader cognitive resources of diverse teams. 
%Furthermore, our lack of results in some areas may be explained with a combination of curvilinear effects, additional moderators,  the presence of a ``Diversity Mind-Set''~\cite{van2005diversity} and even publication bias~\cite{homberg2013top}.

%What is exceedingly clear is that the diversity-effectiveness link is complex and moderated by other factors. However, much progress has been made in this area through the development of integrated models like the CEM. Such models provide a richer understanding of how diversity impacts cognitive and social processes in teams, sometimes in concert and sometimes in opposition. We echo the call by Silveira \& Prikladnicki~\cite{silveira2019systematic}  and Rodríguez-Pérez, Nadri \& Nagappan~\cite{rodriguez2021perceived} to pursue more investigations of this complex link and add to it an emphasis on using established models to do so. Despite the challenges in methodology, analysis, and data gathering, a deeper understanding will likely lead to more effective teamwork and more equal opportunities in an increasingly diverse society.

This study has several implications for future studies of how diversity impacts the effectiveness of software teams. First, the role of task-related and social moderators should be investigated more thoroughly. The \textit{categorization-elaboration model}~\cite{van2004work} provides a valuable framework for such research because it integrates the opposing mechanisms of diversity proposed by \textit{cognitive resource diversity theory} and the \textit{similarity-attraction paradigm}. From a practical viewpoint, such research can also drive the development of training and methods to help teams and organizations to leverage their diversity on all sorts of dimensions, and not limited to gender, age, cultural background, and functional role.
Second, more attention should be paid to the beliefs that teams have about diversity and its effects. Such a ``Diversity Mind-Set''~\cite{van2007work} can act as a powerful moderator by making teams aware of their diversity and how it can expand their experience as a team.
Third, future studies can investigate the role of task interdependence as a moderator of the relationship between psychological safety and relational conflict.
Finally, future research should investigate broader definitions of performance and effectiveness. In this study, we mainly focused on stakeholder satisfaction and team morale. Since effectiveness is a multi-faceted construct~\cite{guzzo1996teams}, we likely missed aspects that are affected by diversity in teams, like speed, quality, or innovativeness.

\section*{Acknowledgment}
The authors thank The Liberators BV and its community of patrons for funding part of this research.

\section{Supplementary Materials}
A replication package for the sample study is available at the following DOI to support Open Science: \url{https://www.doi.org/10.5281/zenodo.10092333} under a CC-BY-NC-SA 4.0 license.

 \section{Responsible disclosure}
Data has been collected and stored according to the policy for research data management of Aalborg University, respecting the total anonymity of informants.

 Christiaan Verwijs has a financial interest in The Liberators BV.

\bibliographystyle{IEEEtran}
\bibliography{bib}

\end{document}